# On the effects of alternative optima in context-specific metabolic model predictions

Semidán Robaina-Estévez*, Zoran Nikoloski

Systems Biology and Mathematical Modeling Group, Max Planck Institute of Molecular Plant Physiology, Potsdam-Golm, Germany. Bioinformatics Group, Institute of Biochemistry and Biology, University of Potsdam, Potsdam-Golm, Germany.

*robaina@mpimp-golm.mpg.de

## Abstract

The integration of experimental data into genome-scale metabolic models can greatly improve flux predictions. This is achieved by restricting predictions to a more realistic context-specific domain, like a particular cell or tissue type. Several computational approaches to integrate data have been proposed—generally obtaining context-specific (sub)models or flux distributions. However, these approaches may lead to a multitude of equally valid but potentially different models or flux distributions, due to possible alternative optima in the underlying optimization problems. Although this issue introduces ambiguity in context-specific predictions, it has not been generally recognized, especially in the case of model reconstructions. In this study, we analyze the impact of alternative optima in four state-of-the-art context-specific data integration approaches, providing both flux distributions and/or metabolic models. To this end, we present three computational methods and apply them to two particular case studies: leaf-specific predictions from the integration of gene expression data in a metabolic model of *Arabidopsis thaliana*, and liver-specific reconstructions derived from a human model with various experimental data sources. The application of these methods allows us to obtain the following results: (*i*) we sample the space of alternative flux distributions in the leaf- and the liver-specific case and quantify the ambiguity of the predictions. In addition, we show how the inclusion of $\ell_1$-regularization during data integration reduces the ambiguity in both cases. (*ii*) We generate sets of alternative leaf- and liver-specific models that are optimal to each one of the evaluated model reconstruction approaches. We demonstrate that alternative models of the same context contain a marked fraction of disparate reactions. Further, we show that a careful balance between model sparsity and metabolic functionality helps in reducing the discrepancies between alternative models. Finally, our findings indicate that alternative optima must be taken into account for rendering the context-specific metabolic model predictions less ambiguous.

## Author Summary

Recent methodological developments have facilitated the integration of high-throughput data into genome-scale models to obtain context-specific metabolic reconstructions. A unique solution to this data integration problem often may not be guaranteed, leading to a multitude of context-specific predictions equally concordant with the integrated data. Yet, little attention has been paid to the alternative optima resulting from the integration of context-specific data. Here we present computational approaches to analyze alternative optima for different context-specific data integration instances. By using these approaches on metabolic reconstructions for the leaf of *Arabidopsis thaliana* and the human liver, we show that the analysis of alternative optima is key to adequately evaluating the specificity of the predictions in particular cellular contexts. While we provide several ways to reduce the ambiguity in the context-specific predictions, our findings indicate that the existence of alternative optimal solutions warrant caution in detailed context-specific analyses of metabolism.

# Introduction

Genome-scale metabolic models (GEMs) have proven instrumental in characterizing the activity of metabolic pathways in different biological scenarios. The activity of all metabolic reactions is specified by the flux distribution, which can be readily inferred from GEMs through the usage of constraint-based approaches (1,2). Such approaches often infer fluxes as solutions to a convex optimization problem in which an objective function is optimized under specified constraints. Two types of constraints can generally be considered: The first is due to the stoichiometry, thermodynamic viability (*i.e.*, if a reaction is irreversible or reversible under normal physiological conditions) and mass-balance conditions. These constraints are included in every constraint-based approach. The second type comprises constraints specific to each approach, and usually reflects the context-specific knowledge or data to be integrated. Flux distributions which satisfy the set of constraints are called feasible. A convex optimization problem is guaranteed to render a unique optimal value (3). However, it is not always guaranteed that there is a unique flux distribution realizing the optimal objective value, leading to alternative optimal flux distributions. Indeed, such a space of alternative optima arises even in the case of flux balance analysis (FBA), as a classical representative of constraint-based approaches (4–9).

Experimental systems biology studies have generated comprehensive atlases of transcript, protein, and metabolite levels from different context, such as: cell types, developmental stages, and environments, across different species from all kingdoms of life (10–15). Analyses of these data sets have already pointed that context-specific differences in the levels of molecular components often affect the activity of metabolic pathways. Additionally GEMs allow constraint-based approaches to integrate such data sets through the so-called gene-protein-reaction rules, which relate metabolic reactions with the enzymes involved and their coding genes (16–19). These approaches address two aims: (*i*) obtaining context-specific flux distribution and (*ii*) determining context-specific GEMs; we refer to the respective approaches as flux- and network-centered, respectively. Alternative optima may also result from the integration of context-specific data. In both settings, the existence of alternative optima leads to ambiguity in context-specific flux distributions and/or network reconstructions, since alternative solutions may substantially differ. This is particularly important in the case of context-specific network reconstructions, where further investigations conducted on a single optimal network could lead to erroneous conclusions.

To our knowledge, only three studies considered the space of alternative optimal solutions arising from flux-centered approaches: The approach termed iMAT (20) proposed a procedure to classify the flux state of reactions into active, inactive or uncertain across the alternative optima space. Another approach, abbreviated as EXAMO (21), later used the set of active reactions obtained from the iMAT alternative optima space as input to the approach referred to as MBA (22), a network-centered method, to reconstruct a context-specific network. Additionally, the Flux Variability Sampling (23) was used to sample the alternative space of flux values that are equidistant to the data integrated. Finally, we note that alternative optimal context-specific models have not been recognized in the case of network-centered approaches, and currently, there is no available method for their analysis.

In the present study, we propose a method to quantify the variability of alternative optimal flux values of a flux-centered approach. Additionally, we quantify the effect in the alternative optima of including an additional constraint in the flux values, minimize the total sum of absolute flux values, which has been proposed to obtain unique solutions in a flux-centered method (24). Furthermore, we investigate, for the first time, the space of alternative optimal context-specific models that arise from several network-centered approaches, and analyze the potential impact on further metabolic predictions and biological conclusions drawn. The study is organized in two parts. The first part is dedicated to explaining the mathematical and computational logic of both (*i*) the context-specific data integration approaches herein evaluated, and (*ii*) the methods that we propose to analyze the respective alternative optima. The second part presents the findings obtained from applying the previously described methods to two particular case studies: a leaf-specific reconstruction from the model plant *Arabidopsis thaliana*, and a human liver reconstruction. This second part serves as an illustration of the impact that alternative optima have in context-specific metabolic reconstructions, and may be followed independently from the first part—which is primary addressed to the specialized reader.

## Results and discussion

### Evaluation of alternative optima from context-specific data integration approaches: Computational methods

In this section, we present the mathematical formulation of the computational methods that we developed to investigate the alternative optima of three selected data integration approaches. In all three cases, we first provide an overview of the approach, which is followed by a description of the method to explore its alternative optima space. We start by a representative of a flux-centered approach—a modified version of RegrEx (25)—and the method that we propose to explore its alternative optima, termed RegrEx Alternative Optima Sampling (RegrEx$_{AOS}$). We then focus on Core Expansion (CorEx), also developed in this study, which we take as representative of a network-centered approach. In addition, we show how the optimization program behind CorEx can be adapted to evaluate not only its alternative optima space, but that of FastCORE (26) and CORDA (27), two state-of-the-art network-centered approaches.

#### Alternative optima in flux-centered approaches: the case of RegrEx
*Background*

Given a GEM and (context-specific) gene or protein expression data, the Regularized metabolic model Extraction (RegrEx) method reconstructs a context-specific metabolic model, along with the corresponding flux distribution. To this end, RegrEx finds a feasible flux distribution that is closest to a given experimental data set, and is, therefore, considered a flux-centered approach.

The original RegrEx approach relied on a regularized least squares optimization in which the Euclidean distance between the given gene expression data vector, *d*, and a feasible flux distribution, *v*, *i.e.*, the squared $\ell_2$ norm of the difference vector $\epsilon = d - v$, was minimized (25). The regularization was implemented by also considering the (weighted) $\ell_1$ norm of *v* in the minimization problem, as a means to select the reactions in the GEM that are most important for a given metabolic context. However, here we used a slightly modified version of RegrEx: Instead of minimizing the sum of square errors, we minimize the sum of absolute errors, *i.e.*, the $\ell_1$ norm of $\epsilon$. Except for this substitution, the modified RegrEx version, called RegrEx$_{LAD}$ (for Least Absolute Deviations), follows the same formulation as the original RegrEx (see S1 Appendix for detailed comparison).

The minimization problem behind RegrEx$_{LAD}$ considers a set of constraints required to handle reversible reactions: In this case, absolute flux values must be taken into account when minimizing the distance to the (non-negative) associated gene expression (*i.e.*, for a reversible reaction $i$, $\epsilon_i = |v_i| - d_i$). This is accomplished by splitting reversible reactions into the forward and backward directions, each constrained to have non-negative flux value, and introducing a vector of binary variables, $x$, to select only one of them during the optimization. Altogether, these particularities are captured in the mixed integer linear program (MILP),

$$v_{opt} = \arg\min \mathbf{w}^T(\epsilon^+ + \epsilon^-) + \lambda \|v\|_1$$
$$\epsilon^+ = [\epsilon^+_{irr}; \epsilon^+_{for}; \epsilon^+_{back}],$$
$$\epsilon^- = [\epsilon^-_{irr}; \epsilon^-_{for}; \epsilon^-_{back}],$$
$$v = [v_{irr}; v_{for}; v_{back}] \in \mathbb{R}^+_0,$$
$$x \in \{0,1\}^n$$

s.t.

1. $S_{ext} v = 0$
2. $v_{irr_i} + (\epsilon^+_{irr} - \epsilon^-_{irr}) = d_{irr}$
3. $v_{for_i} + (\epsilon^+_{for} - \epsilon^-_{for}) + xd_{revRxns} = d_{revRxn}$   $\Big\}$,   $i \in R_D$
4. $v_{rev_i} + (\epsilon^+_{back} - \epsilon^-_{back}) - xd_{revRxn} = 0$                          (OP$_1$).
5. $v_{irr\,min} \leq v_{irr} \leq v_{irr\,max}$
6. $v_{for} + xv_{for\,min} \geq v_{for\,min}$
7. $v_{back} - xv_{rev\,min} \geq 0$
8. $v_{for} + xv_{for\,max} \leq v_{for\,max}$
9. $v_{back} - xv_{rev\,max} \leq 0$

In OP$_1$, the flux distribution, $v$, is partitioned into the sets of irreversible ($v_{irr}$), and reversible reactions proceeding into the forward ($v_{for}$) and backward directions ($v_{back}$), and the (reaction) columns of the stoichiometric matrix, $S_{ext}$, are ordered to match the partition of $v$. In addition, the components of the error vector, $\epsilon_i = \epsilon^+_i - \epsilon^-_i$, $\epsilon^+_i$, $\epsilon^-_i \geq 0$, are split into two non-negative variables, $\epsilon^+_i$, $\epsilon^-_i$, as a way to computationally treat the otherwise required absolute values of the components of $\epsilon$. Thus, the $\ell_1$ norm $\|\epsilon\|_1 = \Sigma_i |\epsilon_i|$ is replaced by $\epsilon^+_i + \epsilon^-_i$ in the objective function. ($\epsilon$ is defined only over the set of reactions with associated data, $R_D$ in OP$_1$). Finally, the $\lambda$ parameter corresponds to the weight of the $\ell_1$ norm in the objective function, and is chosen during the optimization as to maximize the Pearson correlation between data and flux values (25).

The convexity of OP$_1$ guarantees finding the minimum distance between experimental data and a feasible flux distribution that is allowed by the constraints. However, it does not guarantee that the resulting flux distribution is the only feasible one that is optimal with respect to a particular context-specific data. This variability in optimal flux distributions may be attributed to two factors. On the one hand, as mentioned above, not all reactions in a GEM are typically associated to data. In contrast to *data-bounded* reactions, there is a set of *data-orphan* reactions comprising non-enzymatically catalyzed reactions, reactions without gene-protein annotation or without associated data for a particular context. Data-orphan reactions do not contribute to the error norm in the RegrEx$_{LAD}$ objective function, described in OP$_1$, and their flux value can vary as long as $v$ satisfies the imposed constraints and its $\ell_1$ norm is preserved. This situation is depicted in Fig 1, where the search for a flux distribution $v$ that is closest to the data vector, $d$, is carried out in the projection of the flux cone, $F = \{v: Sv = 0, v_{min} \leq v \leq v_{max}\}$, where $d$ resides.

On the other hand, the geometry of *F* may preclude certain reactions to obtain an exact match with the data value, when *d* remains outside the projection of *F*. In this case, a set of flux distributions may be equidistant to *d*, thus generating variability also in the optimal flux value of data-bounded reactions.

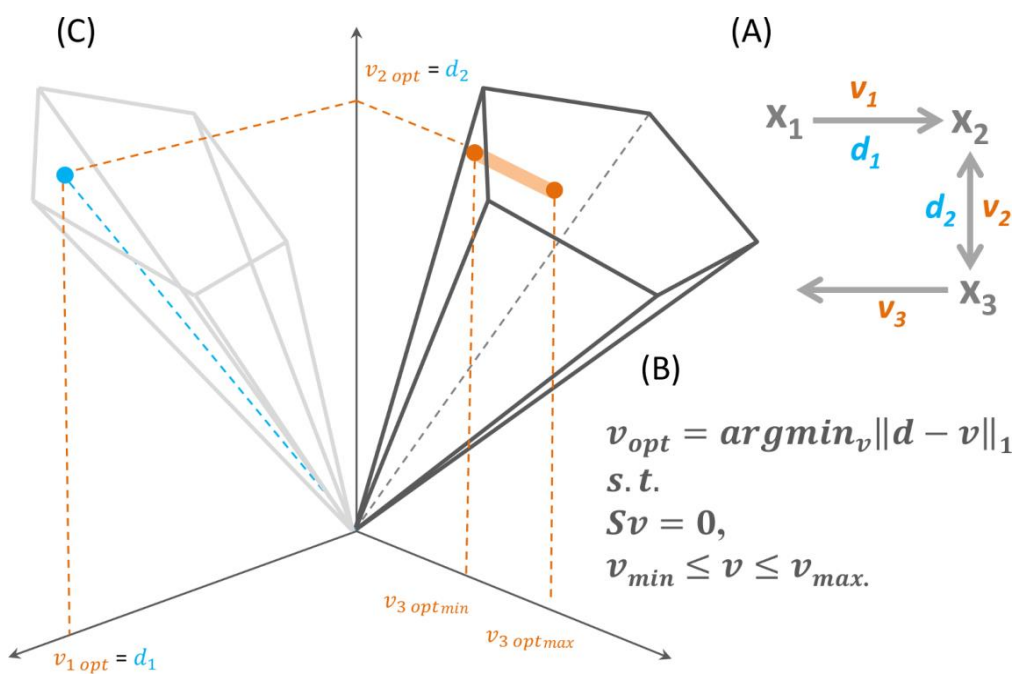

**Fig 1. A depiction of the alternative optima space of a toy RegrEx data integration problem.** (A) A toy data integration problem for a metabolic network with three reactions, $v_{1-3}$, and two reaction-associated data values, $d_{1-2}$ is presented. In RegrEx, the optimization problem consists of finding a flux distribution, $v_{opt}$, which minimizes the distance to the data being integrated and is compatible with the mass balance and thermodynamic constraints. In this example, only two of the three reactions are data-bounded; thus, the third, $v_3$, is free to vary its flux value without affecting the minimum overall distance in (B). This situation is depicted in (C), where the flux cone (the set of flux distributions, $v$, that are compatible with the imposed constraints) is projected onto the two-dimensional space where the data vector, d, resides, and the search for the optimal, $v_{opt}$ is conducted on this projection. This implies that $v_3$ can vary along the direction orthogonal to the projection plane, as long as its value remains within the flux cone (here depicted as the orange line crossing the cone). Hence, the alternative optima space of this data integration problem consists of alternative vectors, $v_{opt(i)}$, in which the components $v_1$ and $v_2$ are fixed, and $v_3$ varies between $v_{3optmin}$ and $v_{3optmax}$.

*The RegrEx alternative optima sampling method*

The general approach followed by RegrEx$_{AOS}$, depicted in Fig 2, is similar to the Flux Variability Sampling (23) (here adapted to RegrEx$_{LAD}$, see S1 Appendix). RegrEx$_{AOS}$ first creates a random flux vector, $v_{rand}$, which is bounded by the maximum and minimum flux values previously calculated by Flux Variability Analysis (using only upper and lower bounds as constraints, see Methods). It then searches for the closest flux vector, $v$, to $v_{rand}$ that belongs to the alternative optima space, *i.e.*, it is at the same distance to the data vector, *d*, and has the same $\ell_1$ norm as the previously calculated RegrEx$_{LAD}$ optimum. This is performed by solving the MILP given in OP$_2$:

$$\min_{\substack{\epsilon^+=[\epsilon^+_{irr};\epsilon^+_{for};\epsilon^+_{back}], \\ \epsilon^-=[\epsilon^-_{irr};\epsilon^-_{for};\epsilon^-_{back}], \\ \delta^+=[\delta^+_{irr};\delta^+_{for}], \\ \delta^-=[\delta^-_{irr};\delta^-_{for}], \\ v=[v_{irr};v_{for};v_{back}]\in\mathbb{R}^+_0, \\ x\in\{0,1\}^n}} \|\delta^+ + \delta^- + \delta_{back}\|_1$$

s.t.

1–9 (OP$_1$)

10. $\epsilon^+ + \epsilon^- = \epsilon^+_{opt} + \epsilon^-_{opt}$

11. $\|v\|_1 = \|v_{opt}\|_1$

12. $v_{irr} - (\delta^+_{irr} - \delta^-_{irr}) = v_{rand(irr)}$

13. $v_{for} - (\delta^+_{for} - \delta^-_{for}) - xv_{rand(revRxn)} = 0$

14. $-v_{back} + \delta_{back} + xv_{rand(revRxn)} = v_{rand(revRxn)}$

(OP$_2$).

Finally, RegrEx$_{AOS}$ iterates this routine *n* times to obtain a sufficiently large sample; here we used *n* = 2000, which is sufficient sample size for the subsequent statistical analyses.

OP$_2$ inherits constraints 1-9 from OP$_1$ and incorporates two sets of new constraints. Constraints 10 and 11 are added to guarantee that *v* renders the same similarity to data and the same $\ell_1$ norm of the previously found RegrEx$_{LAD}$ optimum, $v_{opt}$, respectively. In addition, constraints 12-14 introduce the auxiliary variables $\delta_{irr}$, $\delta_{for}$ and $\delta_{back}$ quantifying the distance of an optimal flux distribution to the randomly generated $v_{rand}$. More specifically, $\delta_{irr(i)} = \delta^+_{irr(i)} - \delta^-_{irr(i)} = v_{rand(i)} - v_{irr(i)}$, $i \in I_R$, acts over the set of irreversible reactions ($I_R$) and $\delta_{for(i)} = \delta^+_{for(i)} - \delta^-_{for(i)} = v_{rand(i)} - v_{for(i)}$, $\delta_{back(i)} = v_{rand(i)} - v_{back(i)}$, $i \in R_R$, over the set of reversible reactions ($R_R$). Note that both $\delta_{irr}$, $\delta_{for}$, are defined as the difference of two non-negative components, which enables us to formulate a linear objective function that renders OP$_2$ computationally tractable. In contrast, $\delta_{back}$ does not require this treatment since it always takes non-negative values (see Fig 2). This is because in OP$_2$, the stoichiometric matrix, *S*, corresponding to the GEM is first modified in the following way: we change the sign of the columns, as well as the entry in $v_{rand}$, corresponding to reversible reactions that were randomly assigned a negative flux value in $v_{rand}$. In this manner, all reversible reactions in $v_{rand}$ operate in forward direction (*i.e.*, are non-negative) which facilitates the optimization process. In addition, $\delta_{for}$ and $\delta_{back}$ are constrained to be mutually exclusive by the same binary variable, *x*, introduced to select only one of the directions in reversible reactions (*i.e.* either forward or backward). In this manner, OP$_2$ will select the direction of reversible reactions that minimizes the overall distance to $v_{rand}$. Finally, reversible reactions whose sign was originally changed in $v_{rand}$ are altered back to their original directions and their sampled flux values are modified accordingly.

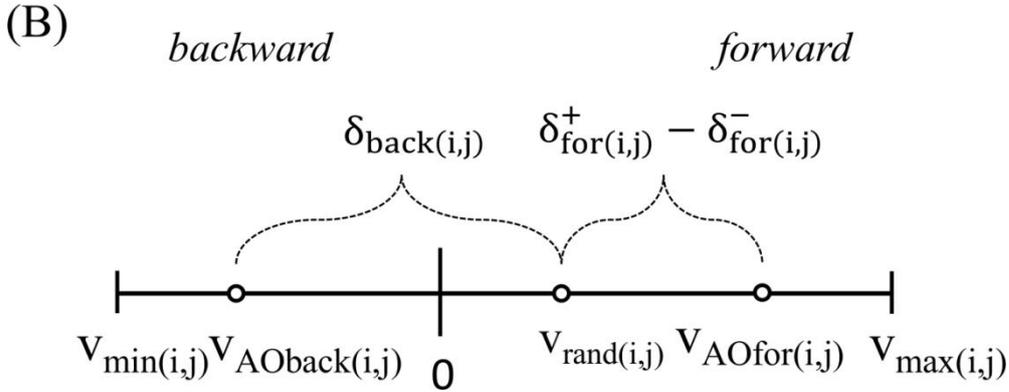

**Fig 2. Pseudocode for RegrEx$_{AOS}$ and details of the treatment of reversible reactions.** (A) RegrEx$_{AOS}$ first finds the minimum, $v_{min}$, and maximum, $v_{max}$, allowable flux values through Flux Variability Analysis (FVA, see Methods) for each reaction in the GEM. It then repeats the following procedure until obtaining the required number of samples (nsamples). (*i*) Generate a random flux distribution, $v_{rand}$, in which each random flux value remains within the feasible range obtained before. (*ii*) Change the sign of the negative entries in $v_{rand}$ and of the corresponding columns in the stoichiometric matrix. (*iii*) Generate an alternative optimal flux distribution, $v_{AO}$, that is closest to $v_{rand}$ through OP$_2$, which takes the modified stoichiometric matrix, $S'$, $v_{min}$, $v_{max}$, $v_{rand}$, the previous optimum RegrEx solution, $v_{opt}$ and the data vector, d, as arguments. (*iv*) Change the sign of the entries in $v_{AO}$ corresponding to the original negative entries in $v_{rand}$. (B) In RegrEx$_{AOS}$, reversible reactions are split into the forward and backward directions. The entries corresponding to reversible reactions in $v_{rand}$ are always non-negative (since the sign is changed if negative), and fall in the range of the corresponding forward direction (since the sign of the associated column in $S$ is changed accordingly). Hence RegrEx$_{AOS}$ can choose between $\delta^+_{for} - \delta^-_{for}$, quantifying the distance between $v_{rand}$ and an optimal flux value in the forward direction, or $\delta_{back}$, which measures the distance between $v_{rand}$ and an optimal flux value in the backward direction. At the end of the optimization process (OP$_2$), RegrEx$_{AOS}$ selects the direction of each reversible reaction that minimizes the overall distance to $v_{rand}$.

# Alternative optimal solutions in network-centered approaches: the case of CorEx

In this section, we analyze the alternative optimal solutions of CorEx, a method that we designed in this study to represent the network-centered approaches. In a general sense, network-centered approaches first partition the set $R = C \cup P$ of reactions in the original GEM into a core set, $C$, that must be present in the final context-specific model, and a non-core set, $P$, which does not necessarily have to be in the final model. These approaches find then a subset $P_A \subseteq P$ of non-core reactions that renders $C$ consistent, *i.e.*, all reactions in the core are able to carry a non-zero flux in at least one steady-state solution. The final context-specific subnetwork is then defined as $R_A = C \cup P_A$. Some approaches, like MBA (22), mCADRE (28) and FastCORE (26), aim at minimizing the size of $P_A$, as to obtain a parsimonious final model. In contrast, CORDA (27) relaxes the parsimony condition as a way to prevent eliminating important reactions for a given context. In this respect, CorEx aims at obtaining a parsimonious model, although, as shown in the following, it can be easily adapted to allow increasing the size of $P_A$ if desired.

CorEx follows the MILP displayed in OP$_3$, which minimizes the number of reactions with non-zero flux in $P$ while constraining all reactions in the core to carry at least a small positive flux ($\epsilon$ in constraints 2-3). This is achieved by minimizing the norm ($Z$ in OP$_3$) of the vector, $x$, of binary variables (constraints 4-7) which selects the set $P_A$ that renders the MILP feasible. Note that the selected non-core reactions are forced to carry a small positive flux (constraints 5, 7) to guarantee that they are active in the final context-specific model. Finally, like in RegrEx, reversible reactions are split into the forward and backward directions, to operate only with non-negative flux values. In addition, another vector of binary variables, $y$ in constraints 8-9 of OP$_3$, is introduced to select the direction of reversible reactions (*i.e.*, imposing $v_{for} > XOR\ v_{back} > 0$, when the reaction is selected to be active).

$$Z = \min_{\substack{v=[v_{irr};v_{for};v_{back}]\in\mathbb{R}_0^{r+},\\ x=[x_{irr};x_{rev}]\in\{0,1\}^P\\ y\in\{0,1\}^{rev}}} \|x\|_1$$

s.t.

1. $S_{ext} v = 0$
2. $v_{irr(i)} \geq \varepsilon$
3. $v_{for(i)} + v_{back(i)} \geq \varepsilon$ $\quad\}, \; i \in C$
4. $v_{irr(i)} - x_{irr(i)} v_{max} \leq 0$
5. $v_{irr(i)} - x_{irr(i)} \varepsilon \geq 0$
6. $(v_{for(i)} + v_{back(i)}) - x_{rev(i)} v_{max} \leq 0$ $\quad\}, \; i \in P$
7. $(v_{for(i)} + v_{back(i)}) - x_{rev(i)} \varepsilon \geq 0$
8. $v_{for} + y v_{max} \leq v_{max}$
9. $v_{back} - y v_{max} \leq 0$

(OP$_3$).

To identify alternative optimal CorEx extracted networks, we developed the MILP displayed in OP$_4$. The general idea behind OP$_4$ is to find the most dissimilar context-specific network, $R_{A^*} = C \cup P_{A^*}$, to a previously found optimal $R_A$, that maintains the set $C$ consistent. Namely, it maximizes the number of differences between the reactions contained in $P_A$ and $P_{A^*}$. Note that OP$_4$ inherits constraints 1-9 from OP$_3$, and incorporates three new constraints. Constraint 10 guarantees that the cardinality of $P_{A^*}$ equals that of the previous optimal $P_A$ in OP$_3$. Constraint 11 introduces two additional binary variables, $\delta^+$, $\delta^-$, which measure the mismatches between the vectors $x$, selecting the reactions in $P_{A^*}$, and the optimal vector $x_{opt}$, selecting the reactions in $P_A$ and previously found by OP$_3$. Finally, constraint 12 is added to impose a $\delta^+$ XOR $\delta^-$ relationship to avoid the trivial optimal solution in which $\delta^+ = \delta^-$,

$$\max_{\substack{v=[v_{irr};v_{for};v_{back}]\in\mathbb{R}_0^{r+},\\ x=[x_{irr};x_{rev}],\delta^+,\delta^-\in\{0,1\}^P\\ y\in\{0,1\}^{rev}}} \|\delta^+ + \delta^-\|_1$$

s.t.

1–9. (OP$_3$)
10. $\|x\|_1 = Z$
11. $x + \delta^+ - \delta^- = x_{opt}$
12. $\delta^+ + \delta^- \leq 1$

(OP$_4$).

However, besides CorEx, OP$_4$ can be used to generate alternative optimal networks to other network-centered approaches. We just need to set $x_{opt}$, in constraint 11, to be the optimal $x$ vector of the particular approach under study; in addition, we need to update $Z$, in constraint 10, to the corresponding number of non-core reactions added by this approach (*i.e.*, the size of $P_A$). Note that $x_{opt}$ can be easily constructed from the set $P_A$, which is derived from a particular context-specific model. In addition, a similar constraint to the constraint 10 of OP$_4$, namely $\|x\|_1 \geq Z_{lb}$, may be included in OP$_3$, as a lower bound to its objective function, where $Z^* \leq Z_{lb} \leq R$, and $Z^*$ is the unconstrained optimum of $OP_3$. It is in this manner that CorEx allows relaxing the parsimony condition, as commented before, although in this study we did not constrain the CorEx optimum.

Noteworthy, the main advantage of using $OP_4$ to obtain alternative optimal networks lies in its MILP formulation. This is because, with the exception of CorEx, which also relies on a single MILP, all existing network-centered approaches require iteratively solving a convex optimization problem. For instance, the linear programs behind the consistency testing step of FastCORE (26), or the ones behind the flux balance analysis, iterated over each reaction of the GEM, in CORDA (27). Alternative optima may arise in each one of these iterations, thus exploring the alternative optima space in each case would require an extensive computational effort. In contrast, we circumvent this problem with $OP_4$ by analyzing the alternative solutions of a single MILP. However, $OP_4$ only generates a single, maximally different, alternative optimal network. To generate a sample of alternative networks, here we applied $OP_4$ in an iterative way. We first used $OP_4$ to obtain a maximally different network to a given optimal context-specific network, and then repeated this process of feeding $OP_4$ with the successively generated alternative networks until no additional one was found. At that point, we randomly perturbed the last network by changing the state (active or inactive) of 1% of the reactions, and repeated this process until no additional network was found (an implementation of the procedure is provided in S1File). We note that with this iterative process, which we term the AltNet procedure, we do not guarantee an exhaustive enumeration of all maximally different alternative networks. However, as shown in the next section, it sufficed to illustrate the variety found across optimal context-specific extracted networks in this study.

Finally, we use the AltNet procedure to analyze the alternative optima space of CorEx, FastCORE and CORDA. In the latter case, however, $OP_4$ had to be slightly modified. The reason for the modification is that CORDA divides the reactions in the GEM into four categories, in contrast to CorEx and FastCORE, where only the core, *C*, and the non-core set, *P*, are considered. Concretely, reactions are separated into three groups based on experimental evidence: reactions with *high* (HC), *medium*, (MC) and *negative* (NC) confidence, and an additional group collecting the remaining reactions (OT) in the GEM, for which experimental evidence is not available. In this case, the group HC corresponds to the core set of reactions (*i.e.,* all reactions in HC must be included in the final model), and the other three groups constitute the non-core set *P*, although reactions in MC are preferentially added over NC and OT reactions. To account for the different reaction groups, we partitioned the vector *x* in $OP_4$ into the sets of MC, NC and OT reactions, and evaluated constraint 10 for each of the three sets. In this manner, we guaranteed that an alternative optimal network contained, besides all HC reactions, the same number of MC, NC and OT reactions than the original CORDA optimum.

# Evaluation of alternative optima from context-specific data integration approaches: Case studies

Here, we illustrate the ambiguity found during the extraction of context-specific flux distributions and metabolic networks due to the alternative optima. To this end, we apply the methods described in the previous section to two case studies: a leaf-specific scenario, the model plant *Arabidopsis thaliana,* and a human, liver-specific reconstruction. In the first case, we used the AraCORE model, which includes the primary metabolism of *Arabidopsis thaliana* (29), and a leaf-specific gene expression data set, obtained from (30) (Methods). In the second case, we employed Recon1, a well-established human metabolic model (31). Moreover, we considered two different core sets of reactions that were identified as liver-specific by experimental evidence (taken from [19] and [20]), and upon which the liver reconstructions were built. In addition, we reduced the original metabolic models by taking only the consistent part of them. The resulting models are termed here Recon1red and AraCOREred, and contain a total number of 2469 and 455 reactions, respectively (see Methods for details).

We first analyzed the alternative optima space of RegrEx$_{LAD}$—as a representative of a flux-centered approach—and evaluated the ability of the $\ell_1$-regularization of RegrEx$_{LAD}$ to reduce this space. To this end, we focused on the leaf-specific scenario; however, we also applied these methods to the liver-specific scenario, to verify if our main conclusions held in the case of a larger genome-scale model. We then applied CorEx, a network-centered representative, to extract and analyze the alternative optima for the leaf- and the liver-specific reconstructions, and compare its performance with that of FastCORE [19], a well-established approach. In addition, we evaluated the alternative optimal liver-specific networks generated by CORDA, a recently published approach [20]. Finally, we also investigated the alternative optima of iMAT to the leaf- and liver-specific scenario with both, the original approach proposed in [16] and our own complementary method.

## Alternative RegrEx$_{LAD}$ optima during leaf-specific data integration

After applying RegrEx$_{LAD}$ with $\lambda = 0$, we obtained an optimal, leaf-specific flux distribution. We then applied RegrEx$_{AOS}$ to evaluate the alternative optima space of the previously obtained optimum. The results from this evaluation confirmed the existence of an alternative optima space for RegrEx$_{LAD}$. However, the variability of the fluxes at the optimal objective value was not uniform across different reactions. As expected, data-orphan reactions exhibited more broadly distributed flux values at the alternative optima than data-bounded reactions. We quantified this property by the Shannon entropy (Methods), as a measure of uncertainty of flux value prediction associated to a data integration problem. In this sense, data-orphan reactions showed a larger mean entropy value of 1.64 in comparison to the value of 0.95 found for the data-bounded reactions (one-sided ranksum test, p-value = $1.95 \times 10^{-5}$). However, we found reactions with particularly low or high entropy values in both sets, data-bounded and data-orphan (S1 Table).

This last observation suggests that reactions with low entropy values may be of special importance under the leaf-specific metabolic state. On the other side, high entropy values suggest that the corresponding reactions could operate more freely in the leaf context. For instance, we found that the majority of transport reactions showed large entropy values, in accord with the fact that most transport reactions are data-orphan. Nevertheless, there were some transport reactions with particularly low entropy values, such as: the *TP/Pi translocator* (reaction index 327 in AraCOREred, $H = 0.07$) interchanging glyceraldehyde 3-phosphate and orthophosphate between the chloroplast and cytoplasm, the *P5C exporter* (index 363, $H = 0.01$) exporting 1-Pyrroine-5-carboxylate from mitochondria to cytoplasm and the *ADP/ATP carrier* (index 320, H = 0.01), interchanging ATP and ADP also between mitochondria and cytoplasm (for a comparison, the highest entropy value in the rank is $H = 2.92$, corresponding to the *Proline uniporter*, see the complete list in S1 Table). Therefore, the leaf data integration constrains these transport reactions to take a small range of different flux values due to the network context in which they operate, since they are not directly bounded by experimental data. This observation is contrasted by the high entropy values that these same three reactions when no experimental data are integrated, *i.e.*, when a similar sampling procedure is performed in which only mass balance and thermodynamic constraints are imposed (Methods). In this case, all three entropy values are markedly larger ($H > 2$, S1 Table).

We next focused on the entropy values of reversible reactions in the AraCOREred model. Reversible reactions in a GEM usually correspond to reactions for which no thermodynamic information is available (leaving aside the set which is known to operate close to equilibrium). Therefore, it would be informative to evaluate whether integrating context-specific experimental data in a GEM could be used to fix the direction of such reactions. Interestingly, we found that a large proportion (75.81%) of the reversible reactions carrying a non-zero flux (including data-orphan) had a fixed direction, either forward or backward, in the alternative optima (Table 1). This finding indicates that, even though there is variation in the flux value of reversible reactions, integration of expression data can determine their direction in a given context. Therefore, the proposed approach and findings provide valuable information on how metabolism could be operating under the particular condition.

**Table 1. Summary of the alternative optima space of RegrEx$_{LAD}$ for two case studies, Leaf and Liver, and four values for the parameter λ.** .

| Leaf | λ = 0 | λ = 0.1 | λ = 0.3 | λ = 0.5 |
|---|---|---|---|---|
| H$_{Data}$ | 73.17 | 71.34 | 81.77 | 65.46 |
| H$_{Orphan}$ | 86.82 | 62.18 | 59.97 | 36.50 |
| H$_{Total}$ | 159.99 | 133.52 | 141.74 | 101.95 |
| $\bar{H}_{Total}$ | 1.23 | 1.03 | 1.09 | 0.78 |
| Fixed$_{Rev}$(%) | 75.81 | 75.81 | 80.95 | 98.18 |
| Liver | λ = 0 | λ = 0.1 | λ = 0.3 | λ = 0.5 |
| H$_{Data}$ | 817.22 | 789.37 | 763.68 | 780.87 |
| H$_{Orphan}$ | 810.79 | 658.66 | 488.31 | 310.21 |
| H$_{Total}$ | 1628.14 | 1448.04 | 1251.99 | 1091.08 |
| $\bar{H}_{Total}$ | 1.20 | 1.07 | 0.92 | 0.80 |
| Fixed$_{Rev}$(%) | 61.78 | 60.31 | 62.41 | 52.09 |

For the analyzed sequence of increasing λ-values, the table includes: The sum of entropy values for the subset of data-bounded, H$_{Data}$, and data-orphan, H$_{Orphan}$, reactions, as well as for all reactions, H$_{Total}$, the mean entropy value across all reactions, $\bar{H}_{Total}$, and the proportion of reversible reactions with fixed direction in the alternative optima sample, Fixed$_{Rev}$.

*Effect of regularization on the alternative optima space*

We next evaluated the RegrEx$_{LAD}$ alternative optima space for a sequence of increasing λ-values. This was motivated to test whether the inclusion of $\ell_1$-regularization, besides imposing sparsity in optimal flux distributions, could also reduce the variability found in individual reaction flux values across the alternative optima space. This property could serve as a way to decrease the uncertainty, as measured by the Shannon entropy, associated to a context-specific data integration problem. To this end, we first applied RegrEx$_{LAD}$ on AraCOREred and the same leaf data set, but using three increasing λ-values ($\lambda_1 = 0.1$, $\lambda_2 = 0.3$ and $\lambda_3 = 0.5$). We then applied RegrEx$_{AOS}$ to sample the alternative optima space of each of the three RegrEx$_{LAD}$ data integrations.

We found that the entropy tended to decrease with increasing λ-values, although the effect was more pronounced for the data-orphan reactions (Table 1, Fig 3). For instance, the sum of entropy values among data-orphan reactions decreased from a value of $H_{Orphan} = 86.82$ for λ = 0, to $H_{Orphan} = 36.50$ with λ = 0.5. In contrast, for the data-bounded reactions, it only decreased from a value of 73.17 with λ = 0 to 65.46 with λ = 0.5, and even led to a transient increase at λ = 0.3 (Table 1, Fig 3). These findings suggest that the inclusion of regularization can reduce the uncertainty associated to a context-specific data integration problem. Naturally, there is a trade-off between decreasing uncertainty and increasing sparsity of the obtained models, since greater λ-values also produce smaller models that may exclude reactions that are relevant to a particular context (SFigure1). However, a mild regularization (λ = 0.1) already had a substantial effect in reducing the uncertainty of the RegrEx$_{LAD}$ data integration in this analysis. Specifically, it decreased the total model entropy, defined as the sum of entropy values over all reactions, by 16.54% (from a value of $H_{Total} = 159.99$ for λ = 0, to $H_{Total} = 133.52$ with λ = 0.1, Table 1).

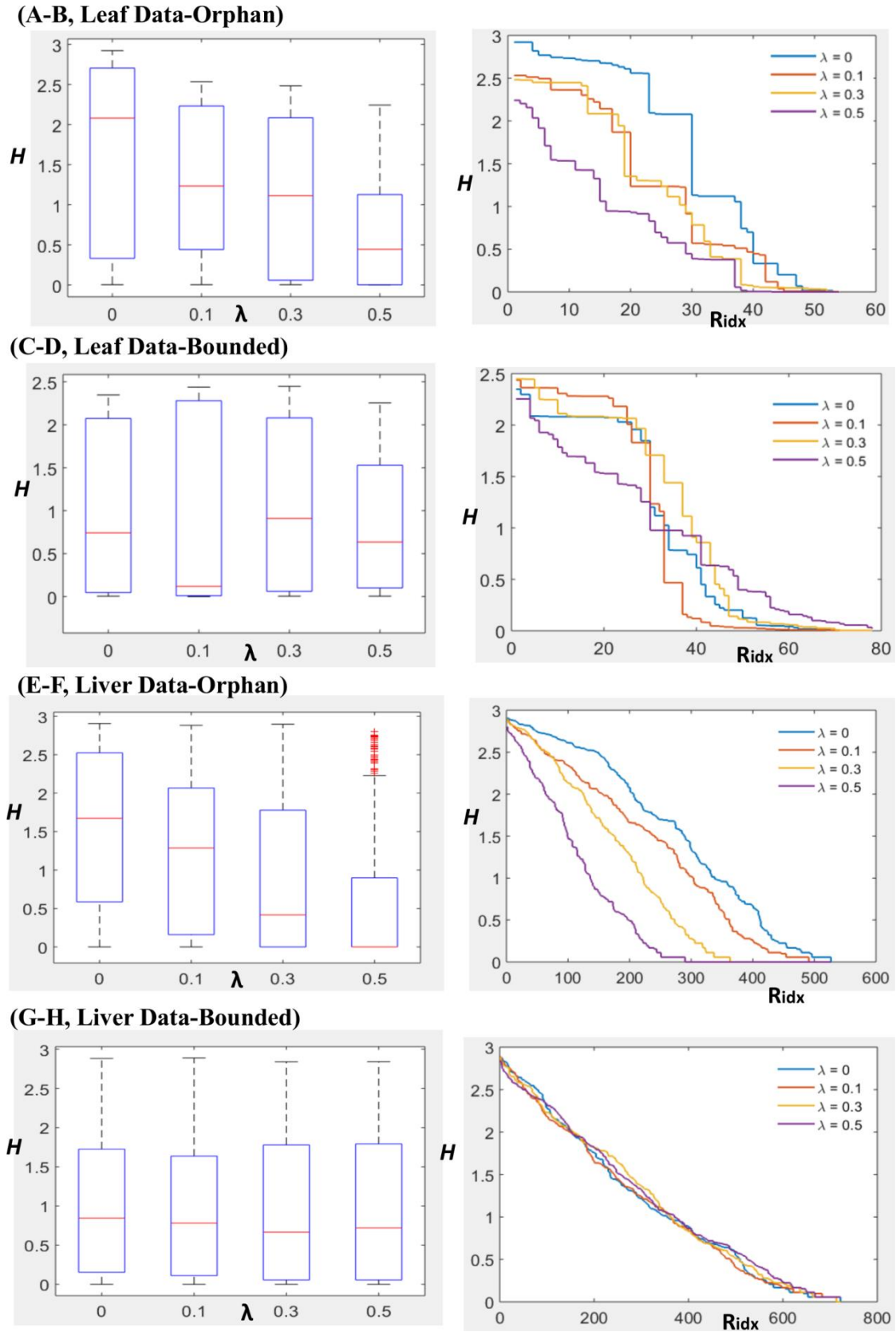

**Fig 3. Effect of regularization on the alternative optima space of RegrEx$_{LAD}$.** The effects of regularization are presented, for the two case studies, by depicting the box plots of the distributions of Shannon entropy values, $H$. The distributions are partitioned into the set of data-orphan (A and E, for leaf and liver, respectively) and data-bounded reactions (C and G, for leaf and liver, respectively) across increasing λ-values. Median values, represented by red lines, decrease monotonically only in data-orphan

reactions (bottom and upper edges in the box plots indicate the 25$^{th}$ and 75$^{th}$ percentile, respectively). Additionally, the individual entropies for each data-orphan (B and F, for leaf and liver, respectively) and data-bounded (D and H, for leaf and liver, respectively) reaction are also presented in decreasing order for the four λ–values (reactions with H < 10$^{-3}$ are omitted). In data-orphan reactions, all distributions with λ > 0 fall below the corresponding to λ = 0 (without regularization, depicted in blue), which is not the case in data-bounded reactions.

Finally, we focused on the effect that regularization had on reversible reactions. We found that the number of reversible reactions with fixed direction increased monotonically with increasing λ-values (Table 1). Hence, this finding suggests that a mild regularization can further constrain the direction in which a reversible reaction can proceed under a particular metabolic context.

*The RegrEx$_{LAD}$ alternative optima in the liver-specific case*

We next analyzed the alternative optima space of RegrEx$_{LAD}$ in the liver scenario. Specifically, we focused on evaluating whether the qualitative results obtained in the leaf context remained unchanged when using Recon1red, a larger genome-scale model. To this end, we used a liver-specific and publicly available gene expression data set (32), and mapped it to the reactions in Recon1red following the same procedure as in the leaf-scenario (Methods). Obtaining samples in a larger model is more challenging, due to the increased computational time required to solve the MILP of OP$_2$. Therefore, we restricted our sample to 100 random points for each of the four λ-values evaluated here, as to avoid an excessively large computational time (the total sample time remained under 41 hours, see Methods for details). In this case, we observed a general qualitative agreement between the leaf and the liver scenarios throughout the increasing λ sequence (Fig. 3, E-H). More specifically, data-orphan reactions showed a monotonic decrease in their median entropy values; however, this effect was less apparent in the case of data-bounded reactions. Specifically, although the total entropy values of data-bounded reactions tended to decreased with increasing λ, with the exception of λ = 0.5 (Table 1), these differences were not significant (one-sided ranksum test, α = 0.05). However, we observed marked differences when looking at the proportion of fixed reversible reactions. In general, this fraction was smaller in the liver scenario, 61.78% *versus* 75.81% with λ = 0 (Table 1), and, in contrast to the leaf case, it did not show an increasing trend with increasing λ-values. We conclude that, while the sample size was smaller than that in the leaf case, these results again suggest that a mild $\ell_1$-regularization of RegrEx$_{LAD}$ can be of help in reducing the ambiguity of context-specific flux values.

## Alternative optima in leaf- and liver-specific metabolic networks

We first applied CorEx and FastCORE to reconstruct two leaf-specific networks, Leaf$_{CorEx}$ and Leaf$_{FastCORE}$. To this end, we used the AraCOREred model and a core set of 91 reactions, which was previously obtained by considering reactions for which the associated gene expression data had a value greater than the 70$^{th}$ percentile (Methods). Both Leaf$_{CorEx}$ and Leaf$_{FastCORE}$ contained the core set and were consistent, *i.e.*, all reactions were unblocked. However, we noticed that Leaf$_{CorEx}$ was more compact than Leaf$_{FastCORE}$, containing 236 versus 254 non-core reactions, respectively (Table 2). We next reconstructed the two liver-specific networks in a similar way. To this end, we used the Recon1red model, and the core set of 1069 reactions defined in the original FastCORE publication (26). In this case, CorEx added 593 non-core reactions to the core set, obtaining the liver-specific reconstruction Liver$_{CorEx}$. FastCORE, on the other hand, added 677 non-core reactions to generate Liver$_{FastCORE}$. Hence, CorEx was able to extract a more compact liver-specific network, resembling the behavior found in the leaf-specific case. After obtaining these context-specific metabolic reconstructions, we searched for alternative optimal networks to all of them, using the AlterNet procedure described in the previous section. To quantify the uncertainty of the leaf- and liver-specific reconstructions, we looked at the number of reaction mismatches between all pairs of alternative networks in each case (computed as the Hamming distance, see Methods). This metric was normalized by the total number of reactions in each metabolic model to allow fair comparison between the two case studies.

**Table 2. Summary of the alternative optima space of the evaluated network-centered methods.**

|  | P | #models | $M_R max$ | $\overline{M_R}$ (CV) | p-value |
|---|---|---|---|---|---|
| **Leaf$_{CorEx}$** | 236 | 61 | 52 [22%] | 29.03(0.29) | 0 |
| **Leaf$_{FastCORE}$** | 254 | 201 | 118 [46.5%] | 66.76(0.54) | |
| **Liver$_{CorEx}$** | 593 | 4 | 156 [26.3 %] | 108.33(0.37) | 0.0022 |
| **Liver$_{FastCORE}$** | 677 | 100 | 398 [58.8%] | 247.93(0.46) | |
| **Liver$_{CORDA}$** | 1527 | 104 | 992 | 545.22(0.42) | 0 |
| **Liver$_{CORDAtest}$** | 1527 | 18 | 860 | 389.40(0.48) | |

This table summarizes the results of the evaluation of the CorEx alternative optima space. It includes the number of added non-core reactions, *P*, the maximum, $M_R max$ (within brackets the percentage of reaction in *P*), and the mean number, $\overline{M_R}$ (CV stands for coefficient of variation), of reaction mismatches (*i.e.,* Hamming distance) across the alternative networks for the leaf- and the liver-specific scenarios evaluated by two methods, CorEx and FastCORE. The last column displays the p-value resulted from a one-sided ranksum test comparing the distributions of Hamming distances between any pair of the alternative networks of CorEx and FastCORE (the null hypothesis states that the distribution generated by CorEx is bigger than that of FastCORE).

We found marked differences between alternative optimal networks in both approaches and metabolic scenarios. In the case of Leaf$_{CorEx}$, alternative networks differed on average in 29 non-core reactions, with a maximum value of 52 reactions (22% of the added non-core reactions). In Leaf$_{FastCORE}$, networks differed on average in 66.78 reactions, and had a maximum number of 118 discrepant reactions (46.5%, Table 2). This situation was even worsened in the liver-specific reconstructions. Between alternative networks to Liver$_{CorEx}$, we found a maximum of 156 discrepant reactions among the 593 in the added non-core (26.3%), with an average of 108.3. In the case of Liver$_{FastCORE}$, the maximum number of discrepant reactions was as high as 398 out of the 677 (58.8%) added non-core reactions, with an average of 246.93 between alternative optimal networks (Table 2).

As a complementary analysis, we also determined the frequency of occurrence of every non-core reaction across the alternative optimal networks. In this manner, we could identify: (*i*) a set of non-core reactions that were always included, termed the active non-core set, (*ii*) a set of non-core reactions that were excluded from all alternative networks, termed the inactive non-core set, and (*iii*) a set of non-core reactions that were included in some of the networks, referred to as the variable non-core set. In this case, we took the size of the variable non-core set as a measurement of the uncertainty of a context-specific network extraction; 28% and a 47% of the total non-core reactions were in the variable set in the cases of Leaf$_{CorEx}$ and Leaf$_{FastCORE}$. On the other hand, a 12% and a 58% were found in Liver$_{CorEx}$ and Liver$_{FastCORE}$, respectively (Fig 4, A-D).

The previous results quantify the structural differences among the generated alternative optimal networks. However, these structural differences do not consider which kind of reactions (*i.e.*, in which pathways in the GEM) are more or less frequent (i.e., ambiguous), in the alternative optima space. To address this issue, we assigned a score (between 0 and 1) to each metabolic pathway based on its representation in the active, variable or inactive non-core set. Specifically, the score represents the fraction of reactions of a given pathway that are assigned to a non-core subset with respect to the total size of the non-core set (Methods). Pathways with high score values in the active and inactive non-core are consistently over- and under-represented, respectively, among the alternative optimal networks. Therefore, these pathways should be more important (the opposite in the inactive non-core case) to maintain the core active and hence the assumed context-specific metabolic function. In contrasts, pathways with high-score values in the variable non-core tend to be represented only in certain alternative optimal networks, thus being more ambiguous in the context-specific reconstruction.

For instance, in the leaf scenario, we found among the pathways with highest score in the active non-core: the *Calvin-Benson cycle, light reactions* and *photorespiration.* All of these pathways showed a maximum score value of 1 in both cases Leaf$_{CorEx}$ and Leaf$_{FastCORE}$, which agrees with key roles of these pathways in a photosynthetic tissue. Additionally, alongside these photosynthetic pathways, we also found housekeeping pathways for the synthesis of AMP, CTP, GMP, UMP, Acetyl-coA or Fatty acid, among others, with the maximum score value in both cases. More interestingly, among the pathways with the highest scores in the variable set we also found primary pathways like the *Tricarboxylic acid cycle, Alanine synthesis,* the *Pentose Phosphate Pathway* and *Pyruvate metabolism*. However, we also found pathways that are usually linked to active photosynthetic tissues like *Starch* and *sucrose degradation* and *sucrose synthesis* (see S9 Table for a complete list containing the ranked pathways).

Moreover, in the liver scenario, we also found typical liver-specific pathways like *Cholesterol Metabolism* and *Fatty acid oxidation* (33) with the maximum score value in the active non-core in the case of Liver$_{CORDA}$. However, we also found a variety of other pathways with high scores in the variable non-core like *CoA catabolism, ROS detoxification* or *Vitamin A metabolism*, which indicates that the variable non-core set contains a diverse set of metabolic functions that may be important to the canonical liver physiology (see S9 Table for a complete list of the ranked metabolic pathways).

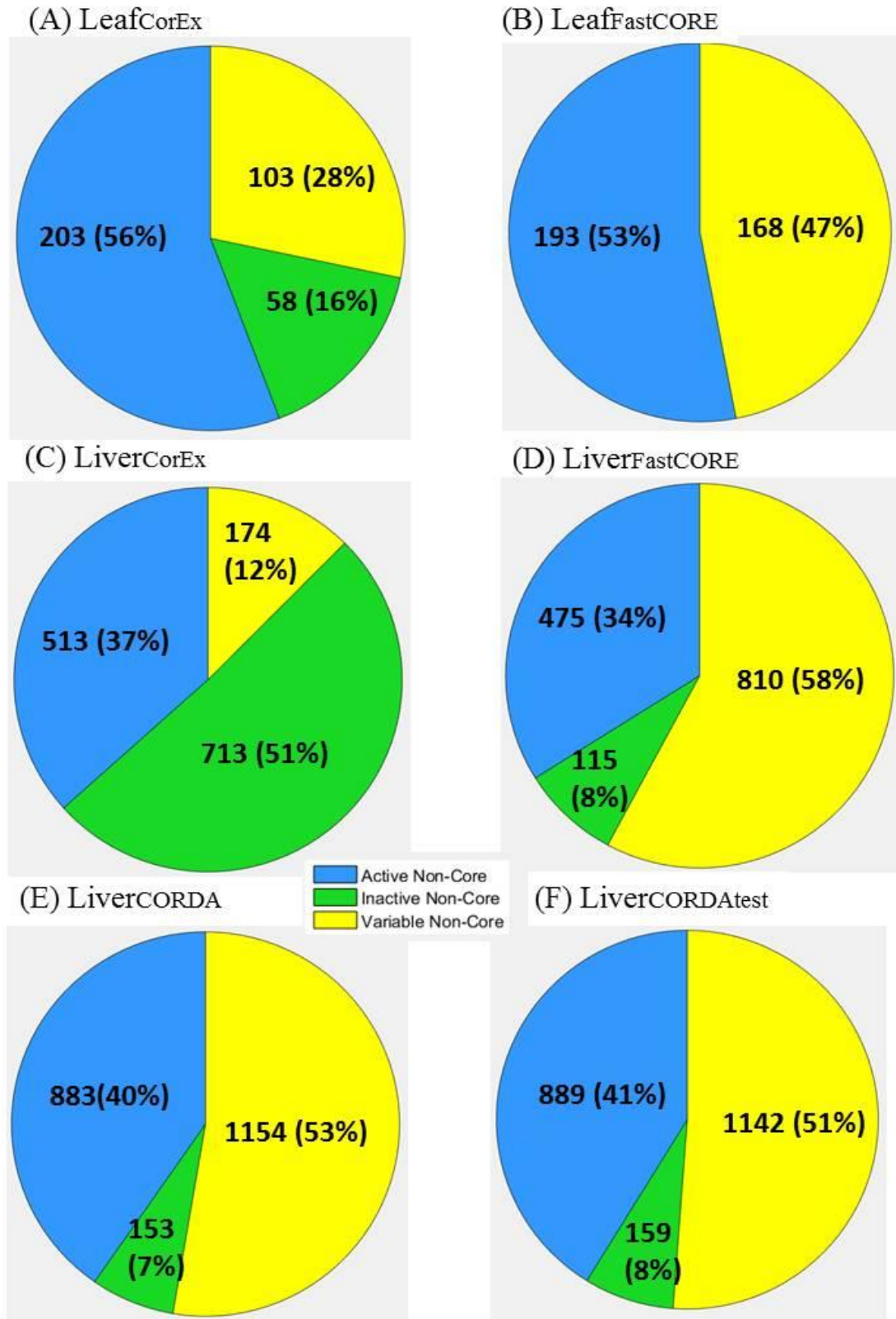

**Fig 4. Alternative optima of CorEx and FastCORE context-specific network extractions.** The results are divided into the leaf-specific scenario for the CorEx (A) and FastCORE (B) alternative optima, and the liver-specific scenario, for CorEx (C), FastCORE (D) and CORDA without applying the metabolic test (E) and applying the metabolic test (F) to further constraint the alternative optima space (see main

text). In all cases, non-core reactions are partitioned into the set that is always included in all alternative networks, (the fixed non-core set, in green), the set that is always excluded (excluded non-core, grey) and the variable non-core set (yellow) which is formed by reactions that are included in some of the alternative networks. In both, the leaf- and the liver-specific scenario, the alternative optima networks generated by CorEx contain a larger proportion of fixed non-core reactions and a smaller proportion of variable non-core reactions. These differences in behavior may be explained by the greater number of non-core reactions that are added by FastCORE, as compare to CorEx, in the optimal solution (see main text).

Finally, we analyzed the alternative optima space of CORDA, a recently published network-centered approach (27). As explained in the previous section (Computational methods) CORDA differs to CorEx and FastCORE in two ways. On one hand, CORDA does not aim at obtaining compact or parsimonious models, but rather emphasizes the metabolic functionality of the final context-specific reconstructions. On the other hand, CORDA considers four groups of reactions based on experimental evidence, out of which only one, the high confidence core set (HC), has to be fully included in the final model (thus being equivalent to the core set of CorEx and FastCORE). In this case, a suitable alternative optimal network must contain not only the entirety of the HC set, but exactly the same number of reactions added by CORDA in each one of the three remaining groups: the medium (MC) and the negative confidence (NC) groups, and the reactions without experimental data (OT). Therefore, it is reasonable to expect that this additional constraint may reduce the uncertainty of the CORDA reconstructions.

To test this idea, we searched for alternative networks to the CORDA liver reconstruction (here Liver$_{CORDA}$) provided in (27). Liver$_{CORDA}$ was obtained from Recon1 and experimental evidence from the Human Protein Atlas (13), and contains 279 HC, 369 MC, 11 NC and 1147 OT reactions. We used again our AltNet procedure, Recon1red (since blocked reactions, by definition, can never be included in a final network), and the classification of the reactions in the four groups also provided in (27). We were indeed able to find alternative networks to the original Liver$_{CORDA}$ with marked differences among them. Concretely, a maximum number of 992 discrepant reactions between two alternative networks, out of the total 1527 distributed among the MC, NC and OT groups (65%, Table 2), with a mean number of 545.22. Similarly, 51% of the non-core reactions (MC, NC and OT) in Recon1red were assigned to the variable non-core set (Fig 4, E).

The examples presented here show that the context-specific reconstructions are more ambiguous than specific, especially in the human liver scenario. This latter case is of special concern, given the implications of obtaining accurate context-specific reconstructions in biomedical research. In fact, most, if not all, of the network-centered approaches have focused on human metabolism (22,26–28). There are ways, however, to cope with this ambiguity or uncertainty of context-specific reconstructions. For instance, as commented before, CORDA aims at obtaining functional reconstructions. In fact, the authors in (27) tested the capability of the Liver$_{CORDA}$ model to conduct a basic set of liver metabolic functions, including aminoacid, sugar and nucleotide metabolism.

We evaluated the alternative Liver$_{CORDA}$ models with the same metabolic test (Methods), and extracted the subset that passed it. Among these networks, we found that the number of discrepancies and the size of the variable non-core were significantly reduced, as compared to the total set of alternative networks (Table 2, Fig 4, E-F). This is not surprising, since requiring the alternative networks to fulfill certain metabolic functions indirectly imposes an additional constraint to the optimal solution. On the other hand, this additional constraint can also be realized by augmenting the core set, as to guarantee that certain key reactions are present in the final context-specific network. This relates to an additional way to reduce the ambiguity of the reconstruction. In the case studies evaluated here, we found that the CorEx alternative networks tended to be more similar among each other than that of FastCORE or CORDA, as quantified by the (normalized by non-core size) number of discrepancies (Table 2). These differences may be explained by the number of non-core reactions included in the optimum: CorEx obtained more compact models than FastCORE in the Leaf- and the Liver-specific case. This imposes a more stringent constraint when searching for alternative optimal networks. However, there is a tradeoff between model parsimony and functionality. In fact, the $Liver_{CorEx}$ model was not able to pass the metabolic function test, while $Liver_{FastCORE}$ was able to pass it. In this particular case, $Liver_{CorEx}$ did not contain the 9 basal exchange reactions (Methods) required to perform the metabolic functions in the test. However, including these 9 reactions in the liver core set sufficed to generate a $Liver_{CorEx}$ model that passed the test.

The analysis of the alternative optima space can be employed to cope with the ambiguity of a context-specific network reconstruction. Notably, the authors of EXAMO (EXploration of Alternative Metabolic Optima) (21) proposed a first step in this direction. In this case, EXAMO first generates a sample of alternative optimal flux distributions of iMAT (20). It then focuses on the activity state of each reaction across the sample, for which it binarizes the flux values through the usage of an arbitrary threshold value. A reaction is included in the *High Frequency Reaction* (HFR) set if it is active throughout the alternative optima sample. Finally, EXAMO uses the HFR set as a core set to MBA (22), a network-centered method, which reconstructs the minimal network that renders the HFR set consistent. EXAMO directly addresses the problem of alternative optima. However, the final context-specific model is again subject to the effects of alternative optima, since a set of alternative networks, all containing the HFR set as a core, could be found for the MBA method.

A possible way to circumvent this problem in the case of iMAT could be the following: *i*) similar to EXAMO, obtain samples of alternative optimal flux distributions, binarize flux values and rank the reactions according to the number of times that they appear as active in the sample, *ii*) include the reactions that are always active (the HFR set) in a core set and the rest in a non-core set, and *iii*), add non-core reactions in decreasing order of frequency until consistency of the core is reached. In this manner, this ranking provides a way to select which non-core reactions are included in the final model. This idea parallels that of mCADRE (28), although in the latter, reactions are ranked following an heuristic approach that considers experimental evidence from several databases, which may be difficult to obtain for certain metabolic contexts. Finally, to generate the sample of alternative optima flux distributions of iMAT, we propose a sampling method similar to $RegrEx_{AOS}$ that allows drawing arbitrarily large samples, as opposed to the one used in EXAMO which generates samples of restricted size. Details about this method, here called $iMAT_{AOS}$, can be found in S2 Appendix.

In the case of the network-centered approaches here evaluated, establishing a ranking of non-core reactions could also be a way to deal with the ambiguity during network reconstructions. Non-core reactions that occur with high frequency in the alternative optima space should be preferentially included in the final network, while reactions with a low frequency should be discarded. To guarantee that the final network is consistent (*i.e.* the core set is active), non-core reactions could be again added in decreasing order of frequency to the core set until consistency is reached. Naturally, this requires the development of competent methods to sample the alternative space of network-centered approaches. In this sense, we consider our proposed AltNet procedure a first step towards this goal.

## Conclusions

We analyzed the space of alternative optima resulting from the integration of context-specific data into GEMs. To this end, we evaluated a representative set from the flux- and network-centered approaches. We selected RegrEx (25) as a representative of flux-centered approaches and CorEx, as a network-centered approach, proposed in this study. In addition, we adapted CorEx to obtain alternative optimal networks for FastCORE (26) and CORDA (27), two state-of-the-art network-centered approaches. We compared the developed approaches and implemented tools on two illustrative case studies: (*i*) a medium size GEM of the primary metabolism of *Arabidopsis thaliana* (29) and a leaf-specific gene expression data set (30), and (*ii*) a larger GEM collecting a reconstruction of a human metabolic network (31), two liver-specific core sets of reactions (26,27) and a liver-specific gene expression data set (32).

Our findings demonstrated the existence of a space of alternative optima for all evaluated approaches integrating context-specific data. Consequently, this space of alternative optima induces ambiguous context-specific reconstructions. In the case of flux-centered approaches, $RegrEx_{LAD}$ in this study, we proposed the usage of a mild regularization to mediate the uncertainty of the resulting context-specific fluxes. In network-centered approaches, our results showed the existence of markedly disparate alternative context-specific networks in CorEx, FastCORE and CORDA. A delicate balance between model parsimony and metabolic functionality seems key to reducing the ambiguity of the context-specific reconstructions. Additionally, an evaluation of the alternative optima space followed by a ranking of the reactions according to their frequency may serve as a way to determine their context-specificity. On this line, we proposed the AltNet procedure to generate alternative optimal context-specific networks.

As a concluding remark, we acknowledge the utility of the existent experimental data integration methods, since they allow a fast and automated generation of context-specific flux distributions and metabolic networks. However, our findings indicated that the interpretation and further usage of their results warrant caution. Specially, since the existence of alternative optima is likely linked to the nature of the context-specific data integration problem, and thus is independent of the approach used. The latter claim is supported by our evaluation across qualitatively different approaches. We advocate the view that an analysis of alternative optimal solutions should be performed, whenever possible, if context-specific data are integrated in metabolic models. In the case of context-specific networks reconstructions, more reliable results could be obtained from subsequent careful knowledge-based curation.

### Methods

This section contains the details about the implementation of the methods described in this study, the GEMs and context-specific data employed in the case examples, and the computation of the distance metric between alternative optimal networks. In addition to this section, the MATLAB code containing the entire workflow followed in this study can be found in the Supplementary Information.

## RegrEx$_{LAD}$, RegrEx$_{AOS}$, CorEx and AltNet implementations

All optimization programs used in this study, (*i.e.,* OP$_{1-6}$) were implemented in MATLAB and solved using Gurobi (version 7.1) (34) on a desktop machine with an Intel Core i7-4790 @3.6 GHz processor and 16GB of RAM. We used default Gurobi parameter values except for: *i)* reduced feasibility tolerance to $10^{-9}$ when solving OP$_{3-4}$, *ii)* increased MIPGap parameter to 1% when solving the MILP of OP$_2$. All generated code with the implementations is available as Supplementary Information.

## Metabolic model and gene expression data

A reduced version of the original AraCORE model (29) was used in this study: AraCORE contains 549 reactions and 407 metabolites assigned to four subcellular compartments, whereas the herein used version (AraCOREred) contains 455 reactions and 374 metabolites. The reactions that were removed correspond to exchange reactions that directly connect organelles to the environment (circumventing the cytoplasm), and were eliminated to avoid bias in the obtained flux distributions. AraCOREred can be found in the Supplementary Material.

Leaf-specific gene expression values were taken from (30), stored in the GEO database under the accession numbers GSM852923, GSM852924 and GSM852925 corresponding to *Arabidopsis thaliana* Col-0 lines with no treatment. The corresponding CEL files were normalized using the RMA (Robust Multi-Array Average) method implemented in the *affy* R package (35). In addition, probe names were mapped to gene names following the workflow described in (36), where probes mapping to more than one gene name are eliminated. Gene expression values were then scaled to the maximum value and mapped to reactions in the AraCOREred model following the included Gene-Protein-Reaction rules and a self-developed MATLAB function, *mapgene2rxn*, which is available in S1File. This process was repeated for the three samples in the dataset and mean values were taken as representative values to obtain the final leaf-specific data used in this study.

Liver-specific gene expression values were obtained from (32), which is accessible under: http://medicalgenomics.org/rna_seq_atlas/download. In this case, we used the RPKM values corresponding to the liver (normal tissues). Since the RPKM values are already normalized we used them directly as input of the *mapgene2rxn* procedure, already described.

We removed blocked reactions from the original Recon1 model to get the Recon1red model used in this study. To this end, we performed a Flux Variability Analysis (see next section) and removed reactions with a maximum absolute flux, $|v_i| < 10^{-6}$. The Flux Variability Analysis was implemented in the MATLAB function *reduceGEM* which also extracted the reduced model, Recon1red, in a COBRA compatible MATLAB structure. The function is available in SFile1.

## Extreme flux values of the flux cone

The minimum and maximum allowed values of each reaction in AraCOREred were determined through Flux Variability Analysis (4). Although only the mass balance and the thermodynamic constraints were imposed (*i.e.,* no reaction was forced to take a fraction of a previously calculated optimal value).This was accomplished through the following linear program,

$$\min_v / \max v_i, \quad \forall i \in v$$

$$s.t.$$

$$Sv = 0$$

$$v_{min} \leq v \leq v_{max},$$

which was implemented in MATLAB and solved with the Gurobi solver (version 6.04). The own-developed MATLAB function can be found in Supplementary Material under the name of *FVA*.

## Sampling flux distributions from the flux cone

To evaluate to what extent the Leaf data integration affected the entropy values of the reactions in the AraCOREred model, we also sampled the space of feasible flux distributions (*i.e.*, the flux cone) when no experimental data was been integrated. To this end, and to allow direct comparability of the results, the flux cone was sampled following a similar approach as in RegrEx$_{AOS}$: first, we generated a random vector of flux values, $v_{rand}$, within the minimum and maximum values obtained by regular Flux Variability Analysis. The closest flux vector $v$ to $v_{rand}$ within the flux cone was then obtained by minimizing the Euclidean distance between the two vectors. The following quadratic program was used to this end:

$$\min_v \tfrac{1}{2} \| v - v_{rand} \|_2^2$$

$$s.t.$$

$$Sv = 0$$

$$v_{min} \leq v \leq v_{max}.$$

This procedure was iterated to obtained a sample of size $n = 2000$. After the sample was generated, we obtained the Shannon entropy values of the samples in the same way as when evaluating the alternative optima space of RegrEx$_{LAD}$ (described in the next section). The MATLAB function implementing this sampling procedure can be found in S1File under the name *coneSampling*.

## Quantification of the RegrEx$_{LAD}$ alternative optima space

The Shannon entropy of the sampled alternative optima distribution, $H_i$, was used to quantify the extent to which the flux values of a reaction, $i$, varied across the alternative optima space. It was calculated as follows:

$$H_i = -\sum_{k=1}^{n} f_{i,k} \log(f_{i,k}).$$

Where $f_{i,k}$ represents the frequency (*i.e.,* number of counts relative to sample size) of the $k$ interval in the distribution, for $n = 20$ equally spaced flux value intervals within the flux range of $i$. In addition, the total entropy of an alternative optima space, $H_T$, was defined as the sum of the entropies corresponding to the $r$ reactions in AraCOREred, *i.e.*,

$$H_T = \sum_{i=1}^{r} H_{v(i)},$$

and was taken as a measure of the total flux variability found in a particular alternative optima space.

## Generation of metabolic networks from context-specific flux distributions and calculation of network distance

In the case of CorEx, we generated the set of alternative optimal metabolic networks from the set of sampled alternative optimal flux distributions. To this end, we first generated the binary vector representations of the flux distributions. The binary vector representations were generated by assigning a value of 1 to the entries corresponding to reactions with a flux value $v \geq 10^{-6}$, and 0 otherwise. This process was repeated for each sampled alternative optimal flux distribution. In addition, repeated vector representations were removed from the generated set. After the binary representations were obtained, we calculated the number of mismatches between any pair, $a,b$, of binary vectors, with $a \neq b$, *i.e.*, the Hamming distance,

$$M_{R(a,b)} = \sum_{k=1}^{n} |a_{(i)} - b_{(i)}|.$$

In this way, we obtained a distribution of $M_R$ values whose characteristics were reported and compared.

## Generation of ranked list of metabolic pathways based on their representation in the active, variable and inactive non-core

We computed a score, ranging between 0 and 1, to quantify the ambiguity found in individual metabolic pathways (subsystems in the GEM) across the space of alternative optimal networks. Concretely, the score of a pathway, $M$, represents the fraction of the reactions in the (total) non-core set, $P$, belonging to the pathway that are assigned to the active, variable or inactive non-core (thus producing a score value for each case). That is, in general,

$$S_X(M) = \frac{X_M}{P},$$

where $X_M \in \{A_M, V_M, I_M\}$ represents the number of reactions assigned to $M$ that are included in the active, variable or inactive non-core, respectively.

## Implementation of the metabolic test applied to the liver-specific reconstructions

We performed the same metabolic test proposed in (27) and applied to the original Liver-specific CORDA reconstruction. This test consists of a list of metabolic tasks that a metabolic model has to perform, including parts of the aminoacid, sugar and nucleotide metabolism. Concretely, there a total of 48 metabolic tasks, divided into the production of different aminoacids from minimal metabolic sources and the excretion on urea (19 tasks), the ability to synthetize glucose from 21 different sources (including some aminoacids), and the production of all 5 nucleotides and nucleotide precursors (8 tasks). The details about these tasks can be found in the original CORDA publication (27), while the MATLAB code of our implementation of this test is provided in S1File. In this study, we used the fraction of performed tasks as measure of the ability of a given liver-specific model to pass this test. For instance, the liver-specific model provided in (27) (under the name of liverCORDAnew), was able to pass 89.58% of the tasks (43 out of 48). In this study, however, we required to pass all tasks in the test to consider an alternative liver-specific network as functional. We realized that, in the liverCORDAnew model, some reactions were slightly different to the analogous reactions in the Recon1red model that we used throughout this study (likely due to different versions of the Recon1 model, which is periodically updated (37)). When we reconstructed our Liver$_{CORDA}$ model, using the same reaction identifiers in liverCORDAnew but extracting the reactions from our Recon1red version, we found that the generated model passed all metabolic 48 tasks in the test. Hence, for consistency of the results, we considered that all proper alternative optimal networks to Liver$_{CORDA}$ had to pass all 48 tasks as well.

# Supporting Information Legends

**S1 Fig. The RegrEx$_{LAD}$ solution path through a sequence of increasing λ-values.** A sequence of optimal solutions (*i.e.*, flux distributions) to the leaf-specific RegrEx$_{LAD}$ integration problem is presented. The sequence begins with λ = 0 (*i.e.,* no regularization) and ends with λ = 1, which is the value for which all fluxes are shrunk to 0. Flux distributions get sparser with increasing values for lambda. In addition, the total entropy of the alternative optima tends to decrease with increasing values for lambda. This indicates the existence of a trade-off between sparsity and entropy reduction. In this study, a mild regularization (λ = 0.1) seems sufficient to substantially reduce the total entropy value while preventing flux distributions to become too sparse (*i.e.*, in which important reactions for a given context may be excluded).

**S1 Table. Ranked list of AraCOREred reactions according to their entropy values across the alternative optima space of RegrEx$_{LAD}$.** In this list, reactions are ranked in descending order according to their entropy values across the alternative optima space of the RegrEx$_{LAD}$ leaf-specific data integration. For each reaction in the list, the reaction index in the AraCOREred model, as well as reaction name, the metabolic subsystem to which it is allocated and the reaction mechanism are displayed. As commented in the main text, the entropy values may be taken as a soft measure of the importance that a reaction has in a given context. This is because small entropy values imply that a reaction is constrained to operate under a small range of flux values in the given context. To contrast the entropy values obtained after the leaf-data integration, we also provide the entropy values shown by reactions when no experimental data are integrated, that is, corresponding to a random sample of the flux cone (see Methods).

**S2 Table. Ranked list of AraCOREred reactions according to their frequency across the alternative optima space of the alternative networks generated by RegrEx$_{LAD}$.**

**S3 Table. Ranked list of AraCOREred reactions according to their frequency across the sample of alternative optimal networks of iMAT.**

**S4 Table. Ranked list of AraCOREred non-core reactions according to their frequency across the sample of alternative optimal networks of CorEx.**

**S5 Table. Ranked list of Recon1red non-core reactions according to their frequency across the sample of alternative optimal networks of CorEx.**

**S6 Table. Ranked list of AraCOREred non-core reactions according to their frequency across the sample of alternative optimal networks of FastCORE.**

**S7 Table. Ranked list of Recon1red non-core reactions according to their frequency across the sample of alternative optimal networks of FastCORE.**

**S8 Table. Ranked list of Recon1red non-HC reactions according to their frequency across the sample of alternative optimal networks of CORDA.**

**S9 Table. Ranked list of metabolic pathways (subsystems) of AraCOREred (leaf) and Recon1red (liver) based on their relative ambiguity in the space of alternative optimal networks of CorEx, FastCORE and CORDA. The description of the score used to rank the pathways is provided in Methods.**

**S1 Appendix. Detailed description of RegrEx$_{LAD}$ and comparison with the original RegrEx$_{OLS}$.**

**S2 Appendix. Description of iMAT$_{AOS}$ and application to the two evaluated case studies.**

**S1 File. MATLAB code containing the implementations of all presented methods as well as the workflow followed to generate all results in this study. The GEM models as well as the (mapped to reaction) expression data used are also included.**